\documentclass[10pt,conference]{IEEEtran}
\usepackage[left=0.7in, right=0.73in, top=0.7in, bottom = 0.98in]{geometry}

\usepackage{graphicx}
\usepackage{setspace}
\usepackage{algorithm} 
\usepackage{algpseudocode}
\usepackage{amsmath,amssymb}
\usepackage{adjustbox} 
\DeclareGraphicsExtensions{.pdf,.jpeg,.png}
\usepackage{enumitem}
\usepackage[noadjust]{cite}

\usepackage{amsmath}

\usepackage{algpseudocode}
\newcommand{\psim}{\boldsymbol{\psi}_m}

\usepackage{amsmath, amssymb}
\usepackage{lipsum}
\usepackage{mathtools}
\usepackage{cuted}
\usepackage{array}
\usepackage{amssymb}
\usepackage{caption}
\usepackage[caption=false,font=footnotesize]{subfig}
\usepackage{multirow}
\usepackage{makecell}
\usepackage{hyperref}

\usepackage{xcolor}
\def\BibTeX{{\rm B\kern-.05em{\sc i\kern-.025em b}\kern-.08em
ThedeltaT\kern-.1667em\lower.7ex\hbox{E}\kern-.125emX}}

\begin{document}

\title{Joint Power Control and Antenna Positioning for Uplink RSMA in Pinching Antenna Systems}

\author{Zar Chi Phyo, Shreya Khisa, Chadi Assi and Sanaa Sharafaddine%
\thanks{Z. Phyo, S. Khisa, and C. Assi are with the CIISE Department, 
Concordia University, Montreal, QC H3G 1M8, Canada 
(e-mail: shreykhasia21@gmail.com, assi@encs.concordia.ca). 
The authors acknowledge the financial support from the Fonds de recherche du Québec (FRQ) 
and from Concordia University (Canada). 
This research is funded by the Fonds de recherche du Québec (FRQ), 
DOI: \href{https://doi.org/10.69777/377241}{10.69777/377241}.}}

\maketitle

\begin{abstract} 
This paper investigates a rate-splitting multiple access (RSMA) for uplink pinching antenna system (PASS). Our objective is to maximize the uplink sum rate by jointly optimizing a continuous antenna positioning and user's transmission power. The formulated problem is highly non-convex and difficult to solve directly; to address this challenge, we propose an alternating optimization (AO) framework which decomposes the original problem into two tractable sub-problems, namely (i) power allocation optimization sub-problem and (ii) antenna position optimization sub-problem. Both sub-problems are solved using successive convex approximation (SCA)-based algorithm and solved alternatively until convergence. The RSMA access for PASS is compared with conventional non-orthogonal multiple access (NOMA) and space-division multiple access (SDMA) techniques. The performance of discrete antenna activation with PASS strategy is also examined. Simulation results demonstrate that our proposed framework significantly enhances the achievable sum rate compared to other multiple access methods.
\end{abstract}

\section{Introduction}

Future wireless networks are expected to become highly digital and intelligent systems, with sixth‑generation (6G) technologies shaping the overall landscape \cite{zhang20196g}. Unlike today’s fifth‑generation (5G) networks, 6G is envisioned to interconnect billions of devices-from self‑driving cars and drones to smart healthcare equipment and diverse sensor ecosystems-demanding very high data rates, ultra‑dense connectivity, user-centric communications, and seamless integration across application domains. Meeting these stringent requirements will rely on advanced antenna architectures and multiple‑access techniques, building on but also extending beyond current solutions. For example, Multiple‑Input Multiple‑Output (MIMO), widely adopted in 5G to improve spectral efficiency and reliability, still faces challenges such as vulnerability to large‑scale fading when Line‑of‑Sight (LoS) links are blocked in complex propagation environments \cite{liu2025pinching}.

Flexible or movable antenna systems have been proposed as a promising way to mitigate the above channel impairments, but they still experience significant large‑scale path loss, particularly at high frequencies \cite{wong2020fluid}. Pinching antenna systems (PASS) have recently emerged as a more flexible alternative \cite{suzuki2022pinching}. A PASS is composed of small dielectric particles, called pinching antennas (PAs), placed along a dielectric waveguide that serves as the transmission medium \cite{liu2025pinching}. By ``pinching” the waveguide with these PA elements, the system can directly radiate signals toward specific users and locations, creating LoS links and thereby reducing large‑scale path loss \cite{ding2025flexible}. Unlike conventional fixed antennas, PASS enables flexible reconfiguration simply by adding, moving, or releasing PA elements along the waveguide, without introducing new radio frequency (RF) front‑end hardware \cite{zhu2023modeling}. These properties make PASS well suited for practical wireless deployments, offering more reliable coverage and enhanced overall system performance.

To further unlock the full potential of 6G, advances in antenna technologies must be paired with next‑generation multiple access (NGMA) schemes-such as non‑orthogonal multiple access (NOMA) and rate‑splitting multiple access (RSMA)-that offer stronger interference management and higher spectral efficiency across diverse network settings \cite{mao2022rate}. Unlike conventional space division multiple access (SDMA), which typically treats interference as noise, or NOMA, which relies on fully decoding interfering signals, RSMA adopts a more flexible strategy by partially decoding interference while treating the remaining part as noise \cite{mao2022rate}. The core principle of RSMA is to split each user’s message into a common stream and a private stream in the downlink, or into multiple sub‑messages in the uplink \cite{mao2022rate}. Combined with successive interference cancellation (SIC), this message‑splitting approach enables RSMA to more effectively manage interference and improve system performance in terms of spectral efficiency, energy efficiency, and user fairness \cite{mao2022rate}.

Several recent studies have investigated PASS-assisted architectures in various settings. In \cite{wang2025antenna}, the authors analyzed a downlink PASS system with multiple waveguides, focusing on antenna activation and resource allocation under NOMA-based access. A low-complexity antenna positioning algorithm for PASS was proposed in \cite{xie2025low}. In parallel, numerous works have examined downlink sum-rate maximization problems with PASS using NOMA and RSMA \cite{xu2025rate,zhou2025sum,wang2025sum}, collectively showing clear performance gains over conventional architectures. In contrast to this rich downlink literature, uplink PASS systems remain comparatively underexplored. The authors of \cite{tegos2025minimum} considered minimum-rate maximization for PASS-based uplink communications with orthogonal multiple access, while \cite{zhang2025uplink} studied multiuser uplink scenarios with PASS. These contributions highlight that uplink PASS faces distinct challenges due to user distribution and interference management \cite{tegos2025minimum,zhang2025uplink}. However, to the best of our knowledge, no prior work has investigated uplink RSMA with PASS, especially from a sum-rate maximization perspective, which directly motivates the present study.

To the best of our knowledge, this work is the first to consider a multi-user uplink PASS using RSMA technology with a single waveguide. We explore a sum rate maximization problem for joint optimization of the antenna’s spatial positions and the users’ transmission power. The formulated optimization problem is highly non-convex with interdependent variables. To address this, we adopt an alternating optimization (AO) approach that decomposes the problem into two sub-problems: antenna positioning sub-problem and power allocation sub-problem. Both of these sub-problems are efficiently solved using the successive convex approximation (SCA) to obtain a high-quality near-optimal performance. Finally, these two sub-problems are solved alternatively until convergence. Simulation results demonstrate that the proposed scheme achieves significant performance improvement over both NOMA and SDMA-based PASS.

\section{System Model}

\subsection{Signal Models}

\begin{figure}
    \centering
\includegraphics[width=0.8\columnwidth]{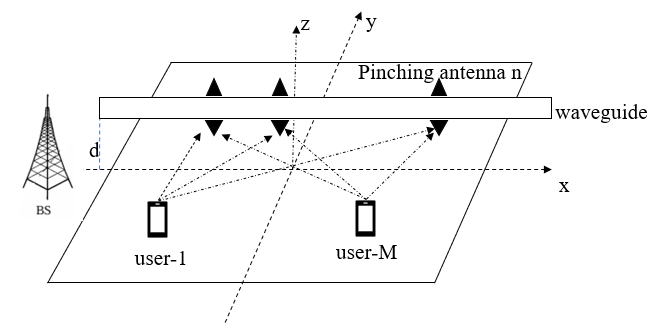}
    \caption{Illustration of uplink PASS}
    \label{fig:Illustration of PA system}
\end{figure}

We consider an uplink PASS communication system (Fig. \ref{fig:Illustration of PA system}) consisting of a BS with $N$ PAs in a set $\mathcal{N} = \{1, . . ,N\}$ located along a single waveguide. We assume that $M$ users in a set $\mathcal{M} = \{1, . . . ,M\}$ are randomly distributed within a rectangular area of a two-dimensional coordination system in the $xy$ plane with side length $D_x$ and $D_y$, respectively. $\psim = (x_m,y_m,0)$ and $\psi_n^P = (x_n^P,0,d)$ denote the position of the $m$-th user in x and y coordinate and the $n$-th PA and $d$ is the height of the waveguide.

In uplink RSMA, it is shown that for $M$ users, splitting the messages of $({M}  - 1)$ users is sufficient to achieve the capacity region \cite{mao2022rate}. Therefore, we assume that the messages of all users except user-$M$ are split. At user $m$, $m \in \{1, .... , {M}  - 1\}$, the message $s_m$ can be divided into $s_{m,j}, \{j =1,2\}$. Particularly, $s_m$ is divided into $s_{m,1}$ and $s_{m,2}$ with power $p_{m,1}$ and $p_{m,2}$ and at user $M$, the message is directly encoded into $s_{{M}}$ with power $p_{{M}}$ \cite{mao2022rate}. Therefore, the transmitted signal of user-$m$ is denoted as,
\begin{equation}
\footnotesize
z_m =
\begin{cases}
\sqrt{p_{m,1}} s_{m,1} + \sqrt{p_{m,2}} s_{m,2}, & \text{if } m \in \{1, \ldots, {M}-1\}, \\[6pt]
\sqrt{p_{M}} s_{M}, & \text{if } m = {M}.
\end{cases}
\end{equation}

The free-space channel between user-$m$ and the PAs is:
\begin{equation} 
\footnotesize
\textbf{h}_m = 
\left[
\frac{\sqrt{\eta} \, e^{-j\frac{2\pi}{\lambda} \|\psim - \psi_1^P\|}}{\|\psim - \psi_1^P\|}, 
\cdots, 
\frac{\sqrt{\eta} \, e^{-j\frac{2\pi}{\lambda} \|\psim - \psi_n^P\|}}{\|\psim - \psi_n^P\|}
\right]^T,
\end{equation}

where $\eta = \frac{c^2}{16{\pi}^2f_c^2}$ \cite{zhang2025uplink}, $c$ is the speed of light, $f_c$ is the carrier frequency and $\lambda$ is the wavelength. Meanwhile, we model the in-waveguide channel as follows,
\begin{equation}
{\textbf{h}_w} = \left[
e^{-j\frac{2\pi}{\lambda_g} \|\psi_0^P - \psi_1^P\|},\ 
\cdots,\ 
e^{-j\frac{2\pi}{\lambda_g} \|\psi_0^P - \psi_n^P\|}
\right]^T,
\end{equation}

where $\psi_0^P$ is the position of the waveguide feed point, and $\lambda_g = \lambda/\eta_{eff}$ is the guided wavelength with $\eta_{eff}$ represents the effective refractive index of the waveguide.

Therefore, the overall channel between user-$m$ and PASS is defined as: 
\begin{equation}
{g}_m = \textbf{h}_{m}^H\textbf{h}_{w}.
\end{equation}

Here $\mathbf{h}_m \in \mathbb{C}^{N \times 1}$ and $\mathbf{h}_w \in \mathbb{C}^{N \times 1}$.

The received signal at the BS can be expressed as,
\begin{equation}
y = \sum_{m \in M}^{} {g}_mz_m + n.
\end{equation}
 ${n}$ is the additive white Gaussian noise (AWGN) with $\mathcal{CN}(0, \sigma^2)$ \cite{zhang2025uplink}.

%
%
\subsection{Rate Analysis}
For uplink RSMA, SIC is performed at the BS. Let the decoding order of sub-messages and messages at the BS be represented by the set $\pi = [\pi_{m,j},\dots, \pi_{M,J},\pi_{M}|\forall j \in J, \forall m \in \mathcal{M}]$ where the elements are arranged in ascending order. Here, $\pi_{m,j}$ denotes the decoding order of the sub-message $s_{m,j}$, while $\pi_M$ corresponds to the decoding orders of the sub-message $s_M$. Specifically, sub-message $s_{m,j}$ is decoded first if its decoding order satisfies $\pi_{m,j} < \pi_{m',j'} $ or $\pi_{m,j} < \pi_{M} $ where $m' \neq m$ and $j \neq j'$, while treating all remaining sub-messages/message as interference. Once $s_{m,j}$ has been successfully decoded, it is removed from the received signal through SIC. The process then continues with the next sub-message according to the decoding order. 

The achievable data rate of user-$m$ and user-$M$ are:
\begin{equation}
\footnotesize
{R}_{m,j} ={\log_2}\left(1+{\frac{{p_{m,j}} |{g}_m|^2 }{\sum_{\pi_{m',j'} > \pi_{m,j}}^{} p_{m',j'}|{g}_{m'}|^2 + \sigma^2}}\right).
\end{equation}

\begin{equation}
\footnotesize
{R}_{M} = \log_2\left(1+\frac{p_{\text{M}}|{g}_M|^2 }{\sum_{ \pi_{m',j'} > \pi_{\text{M}}} p_{m',j'} |{g}_{m'}|^2  + \sigma^2}\right).
\end{equation}
The total achievable rate for user-$m$ is given by the sum of its two sub-messages, as follows,
\begin{equation}
\footnotesize
    R_m=\sum_{j=1}^2R_{m,j},  \quad \mathrm{if} \quad \{m= 1,2, \cdot \cdot \cdot, M-1\}. 
\end{equation}
\section{Problem Formulation}
We aim to maximize the sum rate of the considered uplink PASS by jointly optimizing the transmission power of users and positioning of the antenna. The optimization problem can be formulated as follows,
\begin{equation}
\footnotesize
\begin{aligned}
&\mathcal{P}: \max_{\mathbf{x^p}, \mathbf{p}} \; \left(\sum_{j=1}^2R_{m,j}+R_{{M}}\right),\\
\text{s.t.} \quad 
&\text{C1: } \sum_{j=1}^2 R_{m,j},R_{\text{M}} \geq R_{min}, \forall m \in [1, ..., M-1,M],\\
&\text{C2: } |x_n^P-x_{n'}^P| \geq \Delta, n \neq n', \\
&\text{C3: } \sum_{j=1}^2 p_{m,j}, p_{\text{M}} \leq P_{m}^{max}, \forall m \in [1, ..., M-1,M],\\
&\text{C4: } p_{m,j},p_{\text{M}}  \geq 0,   \forall m \in [1, ..., M-1,M], \notag
\end{aligned}
\end{equation}
where $\mathbf{p} = [p_{11}, p_{12},...,p_{m,j},p_{\text{M}}]$ and $P_{m}^{max}$ is the maximum transmission power of user-$m$ . C1 ensures to meet the requirement of the minimum target rate, while C2 ensures that the distance between consecutive PAs is greater than or equal to $\Delta$ to avoid coupling effects, where $\Delta$ is the minimum inter-distance between two PAs. And, C3 and C4 constrains each user’s transmit power to not exceed its maximum transmit power and should be at least zero. 
%
\section{Solution Approach}
The optimization problem is non-convex due to non-convex objective function and C1. Moreover, the coupling between the optimization variables, transmission power of users and the antenna positioning adds additional complexity, making the problem challenging to solve directly. Therefore, we divide the problem into two sub-problems: for fixed antenna position $\mathbf{x^p}$, we optimize the power allocation $\mathbf{p}$ and for any given power allocation $\mathbf{p}$, the antenna positioning $\mathbf{x^p}$ is optimized. Both of this sub-problems are optimized using SCA. Now, we can reformulate the problem $\mathcal{P}$ as below.
\begin{equation}
\begin{aligned}
&\mathcal{P}_1: \max_{\mathbf{x^P}, \mathbf{p}} \; \sum_{s 
\in S}R_s,  \\
\text{s.t.} \quad 
&\text{C1, C2, C3, C4 }.
\label{eq:opt_problem}
\end{aligned}
\end{equation}
where: \[
\mathcal{S} = \{(m,j)\mid m = 1,\ldots,M-1,\; j = 1,2\} \cup \{M\}.
\].

\newcommand{\xP}{\mathbf{x}^P}
\newcommand{\p}{\mathbf{p}}

\begin{equation}
\footnotesize
R_s
\;=\; \sum_{s\in\mathcal S} \log_2\left(1+\frac{|{g}_s|^2\,p_s}{\;\sum_{s' \neq s}^S|{g}_s'|^2\,p_s' +\sigma^2}\right).
\label{eq:rate}
\end{equation}
where $\mathcal S$ is the set for stream s and $p_s$ is the transmit power of stream s.
\vspace{-10pt}
\subsection{Sub-problem 1: User Transmission Power Optimization}
Given the positions of pinching antennas $\mathbf{x^p}$, the problem \eqref{eq:opt_problem} is rewritten as:
\begin{equation}
\begin{aligned}
&\mathcal{P}_2:\max_{\mathbf{p}} \; R_s, \\
\text{s.t.} \quad 
&\text{C1, C3, C4 }.
\end{aligned}
\label{eq:Rs}
\end{equation}
so that the rate is rewritten as,
\begin{equation}
\footnotesize
  R_s  \;=\;
  \log_2\!\Big(\sum_{s'=1}^S p_{s'}|{g}_{s'}|^2+\sigma^2\,\Big)
  \;-\;
  \hat{R_s},
  \label{eq:Rs_dc}
\end{equation}
where:
\begin{equation}
\footnotesize
  \hat{R_s}  \;=\;
  \log_2\!\Big(\sum_{s'\neq s}^S p_{s'}|{g}_{s'}|^2+\sigma^2\,\Big).
  \label{eq:Rs_hat}
\end{equation}
Since each term in \eqref{eq:Rs_dc} is concave, $R_s$ is a difference of two concave functions and the overall problem in $\mathcal{P}_2$ is non-convex. 
Applying the SCA method to approximate $R_s$ with a concave lower bound in the $t^{th}$ iteration, denote  $\hat{R}_s^{\text{lw}}$ as follows,
\begin{equation}
\footnotesize
  R_s  \; \geq \;
    \log_2\!\Big(\sum_{s'=1}^S p_{s'}|{g}_{s'}|^2+\sigma^2\,\Big)
  \;-\;
  \hat{R}_s^\text{up}  \triangleq R_s^{\text{lw}}
  \label{eq:Rs_dc_2}
\end{equation}
where $\log_2(\cdot)$ is concave and the first-order Taylor expansion of $\hat{R}_s$ at the point $\textbf{p}^t = [p_1^{(t)},...,p_s^{(t)}]$ is defined as $\hat{R}_s^\text{up}$ as follows,
\begin{equation}
\footnotesize
\begin{aligned}
\hat{R}_s^{\text{up}}
&= \log_2\!\Bigg( \sum_{s' \neq s} |{g}_{s'}|^2 p_{s'} + \sigma^2 \Bigg) \\
&\quad + \sum_{s' \neq s}
\frac{p_{s'}|{g}_{s'}|^2}{(\ln 2)\!\left(\sum_{r \neq s} |{g}_r|^2 p_r^{(t)} + \sigma^2\right)}
\big(p_{s'} - p_{s'}^{(t)}\big),
\end{aligned}
\label{eq:taylor_upper}
\end{equation}

Substituting \eqref{eq:Rs} with a concave lower bound in the $t^{th}$ iteration, the equation \eqref{eq:Rs} can be transformed into the following optimization problem:
\begin{align}
 &\mathcal{P}_3: \max_{\mathbf p}\quad\sum_{s\in\mathcal{S}} {R}_s^{\text{lw}}, \notag\\
  \text{s.t.}\quad
  & {R}_s^{\text{lw}} \;\ge\; R_{\min}, s \in\mathcal{S},\label{eq:minrate_split}
\\
  & \text{C3,C4}. \notag \label{eq:power_lin}
\end{align}
Now, we can see that from equations \eqref{eq:minrate_split}, C3, and C4 are convex constraints, and  we are maximizing a concave objective with these constraints, Hence, $\mathcal{P}_3$ becomes a convex optimization problem and can be solved by convex optimization solver such as CVX.
\vspace{-5pt}
\subsection{Sub-problem 2: Antenna Position Optimization}
For a fixed transmit power, $\textbf{p}$, the antenna position vector $\mathbf{x}^{\mathrm P}=[x_1^{\mathrm P},\dots,x_N^{\mathrm P}]^\top$ is optimized to maximize the achievable sum rate, as follows:
\begin{align}
&\mathcal{P}_4:\max_{\mathbf{x^p}} \; R_s,\\
&\quad
\text{s.t.}~\text{C2.}
\label{eq:power_opt_start}
\end{align}
In contrast to power optimization, the channel gain depends on the antenna positions. When the optimization variable $\mathbf{x}^{P}$ is updated at each iteration, the channel gain also varies with antenna positions. Therefore, the rate equation in \eqref{eq:rate} is reformulated with respect to $\mathbf{x}^{P}$ as $R_s(\mathbf{x}^{P})$ and given below:
\begin{align}
\footnotesize
I_s(\mathbf{x}^{P}) 
& = 
\sum_{s' \neq s}^{S} p_s'\,|g_s'(\mathbf{x}^{P})|^2 + \sigma^2, 
\label{eq:interference_def}\\
R_s(\mathbf{x}^{P})
&= 
\log_2\!\big(I_s(\mathbf{x}^{P}) + p_s |g_s(\mathbf{x}^{P})|^2\big)
-\log_2\!\big(I_s(\mathbf{x}^{P})\big).
\label{eq:rate_def}
\end{align}
where $I_s(\mathbf{x}^{P})$ is the denominator of the rate in \eqref{eq:rate}.

The problem becomes a difference of two concave
functions in $\mathbf{x}^{P}$, thus non-convex. To handle this, we use SCA method and apply the first-order Taylor expansion of current iterate as $\mathbf{x}^{P(t)}$.

The squared magnitude of the channel gain term can be written in linearized form as:
\begin{align}
\footnotesize
|g_s(\mathbf{x}^{P})|^2 
\approx |g_s(\mathbf{x}^{P(t)})|^2 
+ \nabla_{\mathbf{x}^{P}}|g_s(\mathbf{x}^{P(t)})|^{2\!\top}
\!\big(\mathbf{x}^{P}-\mathbf{x}^{P(t)}\big),
\label{eq:g_linearization}
\end{align}

Substituting \eqref{eq:g_linearization} into \eqref{eq:interference_def},
we obtain a linearized approximation for the interference term as:
\begin{equation}
\footnotesize
\widetilde I_s(\mathbf{x}^{P})
= I_s^{(t)} 
+ \sum_{s' \neq s}^{S} 
p_s'\,\nabla_{\mathbf{x}^{P}}|g_s'(\mathbf{x}^{P(t)})\big|^{2\!\!\top}
\!\big(\mathbf{x}^{P}-\mathbf{x}^{P(t)}\big),
\label{eq:I_hat}
\end{equation}
and the first term of \eqref{eq:rate_def} is defined as:
\begin{align}
\widetilde H_s(\mathbf{x}^{P})
= \widetilde I_s(\mathbf{x}^{P})
+ p_s\,\widetilde{|g_s|^2}(\mathbf{x}^{P}).
\label{eq:A_hat}
\end{align}

Applying the first-order Taylor expansion to the latter part of \eqref{eq:rate_def}, and a concave lower bound is written as:
\begin{align}
\footnotesize
&{R}_s^{\text{lw}}(\mathbf{x}^{P})
\triangleq 
\log_2\!\big(\widetilde H_s(\mathbf{x}^{P})\big) \notag\\
&-\Big[
\log_2 I_s^{(t)}
+\frac{1}{\ln2\,I_s^{(t)}} (\widetilde I_s(\mathbf{x}^{P}) - I_s^{(t)})\Big] \notag
\\
&\le R_s(\mathbf{x}^{P}).
\label{eq:R_tilde}
\end{align}
At each iteration for given point $\mathbf{x}^{P}$, a concave lower bound of the problem is reformulated as
\begin{align}
&\mathcal{P}_5:\max_{\mathbf{x}^{P}}\quad 
\sum_{s\in\mathcal{S}} {R}_s^{\text{lw}}(\mathbf{x}^{P}) 
 \notag\\
&\text{s.t.}\quad
 \text{C2}. 
\label{eq:SCA_c1}
\end{align}
This problem is convex, as $\mathcal{P}_5$ is a concave maximization subject to convex constraints. It can therefore be solved efficiently by CVX. Furthermore, the overall process, which iteratively solves the two sub-problems is
summarized in Algorithm 1. The objective functions $\mathcal{P}_3$ and $\mathcal{P}_5$ are monotonically increasing functions after each SCA iteration, therefore, the proposed Algorithm 1 is guaranteed to converge.

\begin{algorithm}[H]
\footnotesize
\caption{AO Framework: SCA-User Power \& SCA-Antenna Position}
\begin{algorithmic}[1]
\State \textbf{Inputs:} channels  $g_s, \forall s \in \mathcal{S}$, target rate $R_{\min}$, power budgets $P^{\max}_m, \forall m \in M$, spacing $\Delta$, $\epsilon$.
\State \textbf{Initialize:}  set $t\gets0$, $\mathbf{x^{(0)}_{P}}$ and $\mathbf{p^{(0)}}$; 
\Repeat
  \State \textbf{User Power Optimization:}
    \Statex \quad At fixed $\mathbf{x}^{(t)}_P$,
    \Statex \quad solve the sub-problem $\mathcal{P}_3$ to obtain $\mathbf{p}^{(t+1)}$

  \State \textbf{Antenna Position Optimization:}
    \Statex \quad Using $\mathbf{p}^{(t+1)}$,
    \Statex \quad solve the sub-problem $\mathcal{P}_5$ to obtain $\mathbf{x}^{(t+1)}_P$
    
  \State Get updated sum rate $t \gets t+1$
\Until{$R_s^{(t)} - R_s^{(t-1)} < \varepsilon$}
\State \textbf{Output:} Optimal values : $(\mathbf{p^{\star}}, \mathbf{x^{\star}_{P}})$
\end{algorithmic}
\end{algorithm}

\subsubsection{Initialization  Process}
Since the joint optimization problem is a non-convex problem, the initialization is highly influenced for convergence behavior and final output. Therefore, we initialize the user's transmit powers uniformly within each user’s available power budget. For the position of PAs, the initial layout is selected by uniformly distributing the $N$ antennas along the waveguide. 
\subsubsection{Convergence \& Complexity}

The proposed algorithm is alternatively solved until convergence according to Algorithm 1. First, the power is optimized using SCA where first Taylor approximation is applied to give feasible solution for next iteration. Then, the antenna position is optimized by same approach as power optimization using SCA. The rate is evaluated after each iteration and the best values are stored as the rate increases. This AO framework guarantees that the achievable sum rate satisfies the minimum rate constraints and the algorithm ensures tor the monotonic improvement of the objective function. Let the number of iteration for SCA for the sub-problem-1 and sub-problem-2 are  $K_{\mathrm{SCA}}^{(p)}$ and $K_{\mathrm{SCA}}^{(x_p)}$ respectively. Hence, the complexity of sub-problem-1 is  $O(K_{\mathrm{SCA}}^{(p)} M^{3})$ and the complexity of second sub-problem is $O(K_{\mathrm{SCA}}^{(x_p)} N^{3})$ \cite{tegos2025minimum} The optimization variables are updated iteratively for $K_{\mathrm{AO}}$ iterations. Therefore, the total computational complexity of the proposed algorithm is defined as $O\!\left( K_{\mathrm{AO}} ( K_{\mathrm{SCA}}^{(p)} M^{3} + K_{\mathrm{SCA}}^{(x^p)} N^{3} )\right)$.
\section{Numerical analysis}
In this section, we validate our proposed scheme through numerical analysis. We assume an area of  $D_x \times D_y = 80 \times 80 m^2$ where the users are randomly distributed and strong LoS exists between the users and the PASS. 
The simulation parameters are defined as $M = 2$, $N=6, f_c = 28$GHz, $R_{min} = 0.7$ bps/Hz, $P_{max}=23$ dBm, $ \sigma^2$ = -90 dBm, $n_{eff} = 1.4$ and $d=3$m. We perform Monte Carlo simulations averaging over 50 channel realizations, and to validate the effectiveness of the proposed AO framework, comparisons are made with several benchmark solutions for two settings. The first setting corresponds to continuous antenna positioning for three schemes using Fmincon 1) RSMA-PASS 2) NOMA-PASS 3) SDMA-PASS. The second setting represents (i) discrete activation (selective): a number of antennas are selected to activate and (ii) discrete activation (full): all PAs are activated in discrete pre-configured positions for transmission of signal. 
\begin{figure*}[!t]
\centering
\subfloat[]{%
    \includegraphics[width=0.32\textwidth]{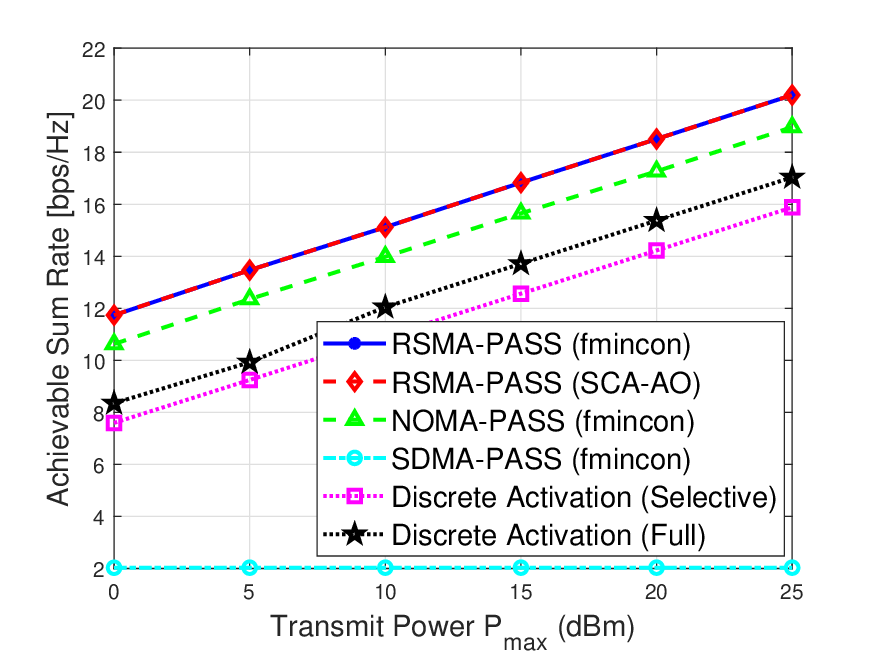}%
    \label{fig:2a}}
\hfill
\subfloat[]{%
    \includegraphics[width=0.32\textwidth]{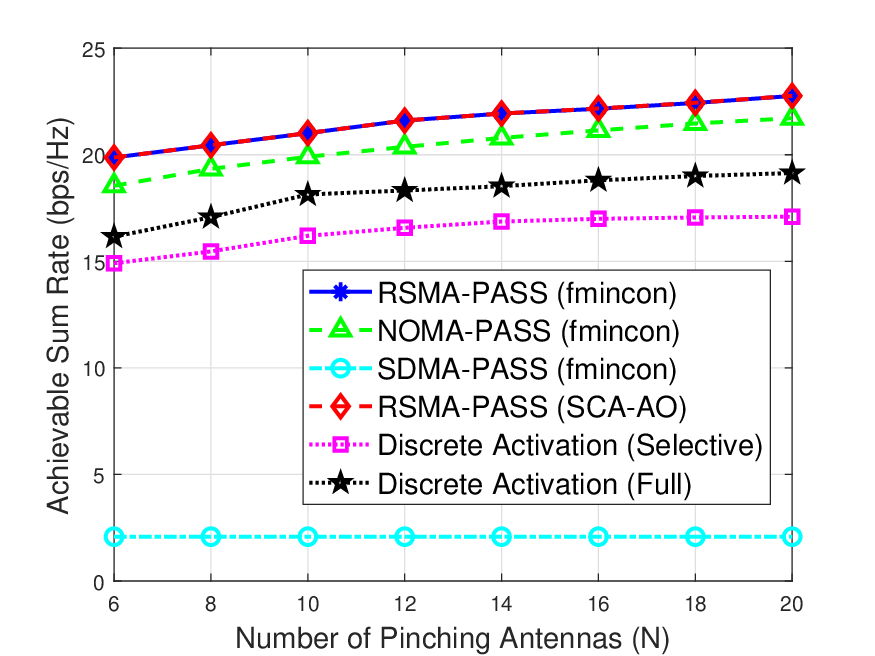}%
    \label{fig:2b}}
\hfill
\subfloat[]{%
    \includegraphics[width=0.32\textwidth]{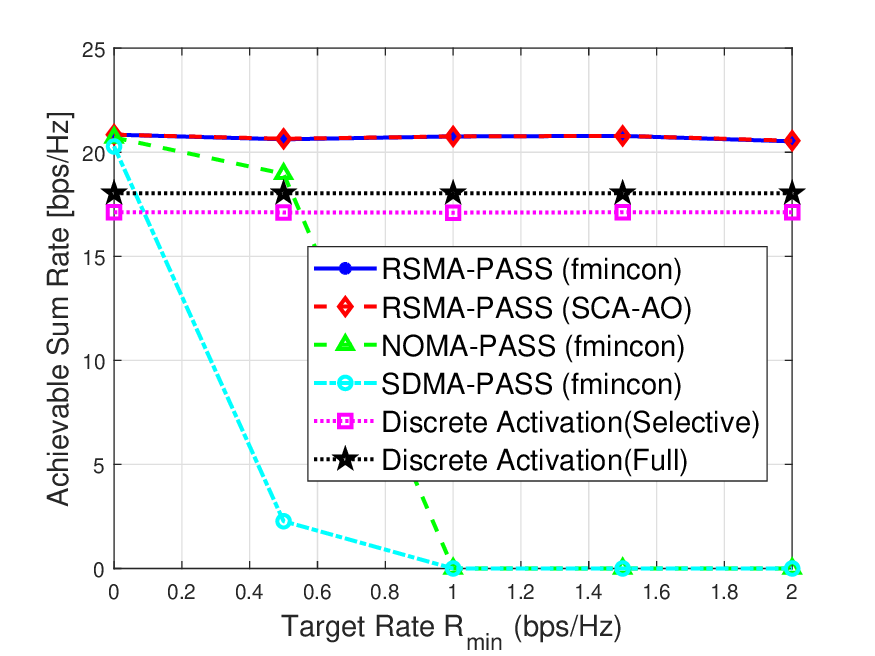}%
    \label{fig:2c}}

\caption{Achievable sum rate for  $M=2$ while varying (a) user transmit power, (b) number of PAs, and (c) target rate $R_{\min}$}
\label{fig:4}
\end{figure*}

\begin{figure*}[!t]
\centering

\subfloat[]{%
    \includegraphics[width=0.32\textwidth]{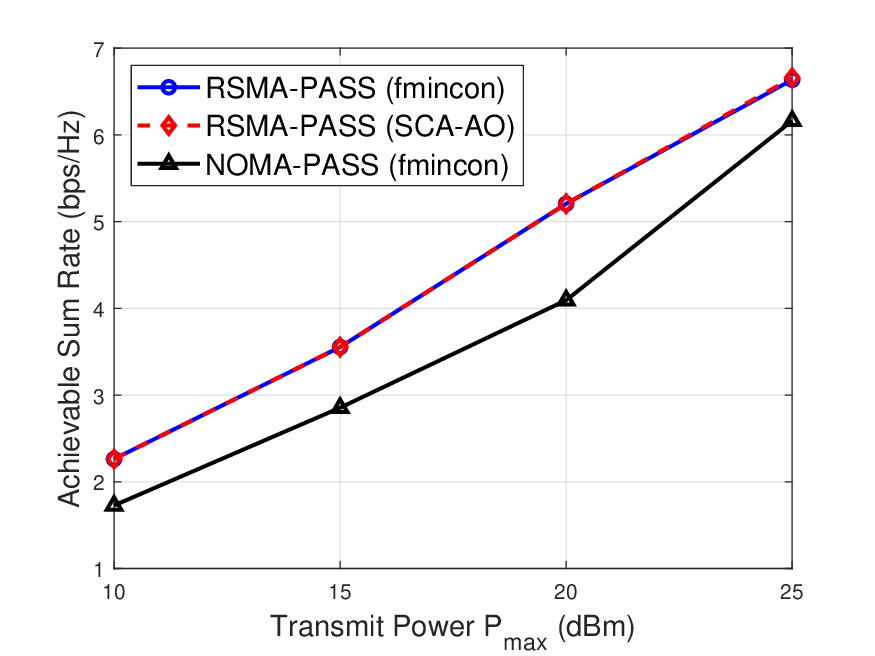}%
    \label{fig:3a}}
\hfill
\subfloat[]{%
    \includegraphics[width=0.32\textwidth]{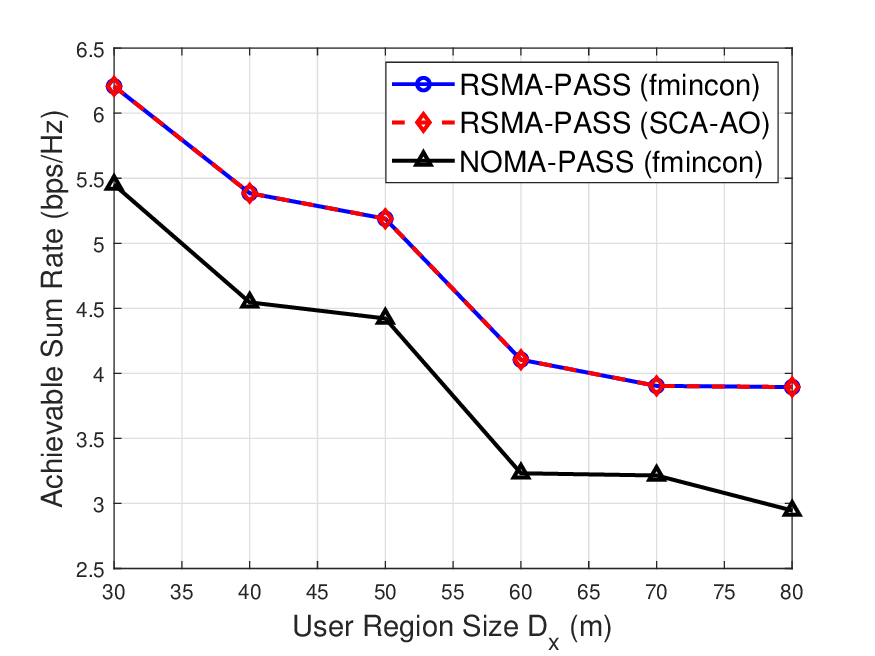}%
    \label{fig:3b}}
\hfill
\subfloat[]{%
    \includegraphics[width=0.32\textwidth]{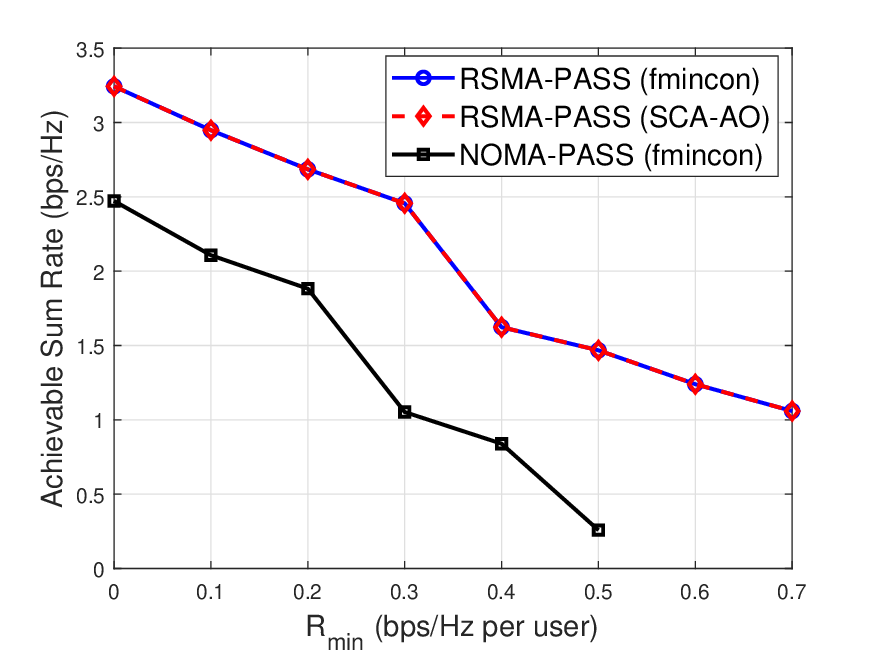}%
    \label{fig:3c}}

\caption{Achievable sum rate for $M=3$ while varying (a) user transmit power, (b) user region, and (c) target rate} 
\label{fig3}
\end{figure*}

Fig.\ref{fig:2a} illustrates the impact of the transmission power of users on the achievable sum rate. The results indicate that the sum rate is monotonically increasing with respect to the transmit power. The proposed RSMA with SCA-AO performs nearly identical to the Fmincon (RSMA) baseline, which validates the verification of the proposed algorithm \footnote{Fmincon in MATLAB is an optimization function for constrained nonlinear optimization problems.}. It can be seen from the figure that the RSMA access outperforms NOMA and SDMA. RSMA achieves approximately 5.3\% over NOMA and 90\% gain over SDMA, respectively. 
This shows that in an uplink PASS aided system, RSMA efficiently manages inter-user interference, while SDMA suffers from strong inter-user interference at high power levels and high minimum rate constraints. Moreover, compared with two discrete activation scenarios, the continuous positioning scheme achieves the highest sum rate because of its ability to dynamically reposition the PAs along the waveguide. In selective discrete activation case, only the subset of antennas are activated, hence, it cannot fully exploit the spatial degrees of freedom, resulting in less sum rate. Overall, the continuous PA positioning confirms the effectiveness of flexible antenna placement in achieving superior performance, whereas discrete activation can offer a practical low-complexity alternative with only moderate performance. 

Fig.\ref{fig:2b} plots the sum rate by varying the number of antennas within the user region when $D= 60m$. Overall, all schemes show improvement in sum rate as the number of antennas increases. This result demonstrates that deploying PAs closer to users can enhance the channel quality which reduces the path-loss, and hence improve the sum rate performance compared to other schemes.

Fig.\ref{fig:2c} shows the impact of varying the target rate $\text{R}_{min}$ from 0 to 2 bps/Hz on the sum rate. As $\text{R}_{min}$ increases, the sum rate of all schemes decreases slightly as expected due to the strict rate requirements. All RSMA schemes are stable because RSMA can meet the strict rate requirements without large performance loss. In contrast, NOMA and SDMA exhibit rapid degradation in performance. For NOMA, when $\text{R}_{min}$ increases, the optimizer gives an infeasible solution because the system does not satisfy it. Meanwhile, SDMA fails to find feasible solutions after the value of $\text{R}_{min}$ reaches 0.8bps/Hz. This result highlights its sensitivity to the minimum rate constraints.  

The simulation parameters for Fig.\ref{fig3} are defined as $R_{min} = 0.01$ bps/Hz and the region size is $D_x \times D_y = 30 \times 30 m^2$. Fig.\ref{fig:3a} shows as user transmit power increases, the achievable sum rate of both schemes increases. It can be seen that RSMA can outperform NOMA in all power levels because RSMA can manage inter-user interference through message, splitting while NOMA is sensitive to SIC ordering in multiuser scenario.
Fig.\ref{fig:3b} illustrates the sum rate by varying the size of the coverage region along $D_x$ while keeping $D_y$ fixed at 30m. It can be seen that when the region size increases, NOMA experiences a lower gain compared to RSMA because its decoding structure struggles under channel disparity and distributed user locations.
Fig.\ref{fig:3c} presents the sum rate while varying the UE target rate within the region of $80m \times 80m^2$. It can be seen that when the target rate is increased, NOMA consistently achieves less performance than RSMA. This is because when the rate constraints increase, NOMA has limited ability to handle high inter-user interference and has strong SIC dependencies, which leads to NOMA performance decreasing by 47\%. In contrast, RSMA with PA system has flexibility in interference management for its message splitting method and adds up with the repositioning capabilities of the PA, RSMA achieves the superior performance in all scenarios.
 
\section{Conclusion}
We proposed an PASS-assisted uplink RSMA framework. We aimed to maximize the sum rate by jointly optimizing antenna positioning and user's transmission power. The formulated optimization problem is highly non-convex in nature and hence difficult to solve directly. Therefore, we decomposed the problem into two sub-problems and leveraged an AO algorithm to solve these sub-problems alternatively. The simulation results demonstrate that our proposed framework can obtain better performance gains over other access techniques such as NOMA and SDMA. In future work, we aim to extend this model into more practical scenarios by considering multi waveguide scenario, and joint decoding order optimization.


\bibliographystyle{IEEEtran}
\bibliography{bibliography}
\vfill
\end{document}